# SANS STUDY OF THE UNILAMELLAR DMPC VESICLES. THE FLUCTUATION MODEL OF LIPID BILAYER.

M.A.Kiselev[1], E.V.Zemlyanaya[2], V.K.Aswal[3]

[1]*Frank Laboratory of Neutron Physics, JINR, Dubna 141980, Russia*
[2]*Laboratory of Information Technologies, JINR, Dubna 141980, Russia*
[3]*Paul Scherrer Institute, CH-5232, Villigen, Switzerland*

On the basis of the separated form-factors model, parameters of the polydispersed unilamellar DMPC vesicle population are analyzed. The neutron scattering length density across the membrane is simulated on the basis of fluctuated model of lipid bilayer. The hydration of vesicle is described by sigmoid distribution function of the water molecules. The results of fitting of the experimental data obtained at the small angle spectrometer SANS-I, PSI (Switzerland) are: average vesicle radius $272\pm0.4$Å, polydispersity of the radius 27%, membrane thickness $50.6\pm0.8$Å, thickness of hydrocarbon chain region $21.4\pm2.8$Å, number of water molecules located per lipid molecule $13\pm1$, and DMPC surface area $59\pm2$Å$^2$. The calculated water distribution function across the bilayer directly explains why lipid membrane is easy penetrated by water molecules.

## INTRODUCTION

Research into the structure of phospholipids, the main component of biological membranes, is very important from the viewpoint of structural biology, chemistry and pharmacology. During the last three years SANS and SAXS experiments were performed to study the structure of unilamellar vesicles. Average radius of the vesicle population, vesicle polydispersity, membrane thickness, and internal structure of membrane were defined on the basis of the hollow sphere model (HS) and the Kratky Porod method [1-4].

The methods developed on the basis of the hollow sphere model have principal limitation: the neutron scattering length density distribution across the bilayer is described by a constant or a strip function. The uniform and the strip-function scattering length density can be considered as zero- and first-order approximation in the evaluation of the internal membrane structure from SANS. The real scattering density profile from neutron diffraction experiments demonstrates a more complex and smooth distribution [5].

In [6], the separated form-factor model (SFF) was developed. It allows one to apply any integrable function to simulate the scattering length density of neutrons across the bilayer for the calculation of coherent macroscopic scattering cross section of vesicle population. The structure factor of the vesicle population was studied via small-angle X-ray scattering in the region of lipid concentration 1-5% (w/w) [7].

In [8], the fitting codes were developed to analyze the structure of the polydispersed population of the unilamellar vesicles on the basis of the SFF model. The parameters of the DMPC vesicle population (average radius, polydispersity, membrane thickness, thickness of hydrophobic part, and number of linear distributed water molecules in the membrane bilayer) were restored only from the SANS spectra, without additional methods (light scattering, diffraction, etc.). The water distribution in the hydrophilic part of membrane was simulated by linear functions of two types: linear function with a break point on the boundary between membrane and bulk solvent and continuous linear function at this boundary. It was shown that continuous function better describes the experimental curve. This analysis was based on the experiment carried out at the YuMO small-angle scattering spectrometer in JINR, Dubna. The SANS curve from the DMPC vesicles at T=30$^o$C was collected in the range of the scattering vector q from 0.008Å$^{-1}$ to 0.2Å$^{-1}$.

The same experiment was fitted in the range of q from 0.037Å$^{-1}$ to 0.14Å$^{-1}$ in the recent paper [9]. The Kratky-Porod method and the 5-strip model of membrane scattering length density were used in [9]. Internal membrane structure of DMPC bilayer and its hydration calculated in [8] and [9] are in sufficient agreement between each other.

An important question is the dependence of evaluated parameters on the width of q range measured in the experiment. To clarify this problem, the same model as in [8], was used in our paper [10] for the evaluation of DMPC vesicle parameters measured at T=30$^o$C at the SANS-1 spectrometer of Paul Scherrer Institute, Swizerland. This curve was collected in the larger range of the scattering vector q: from 0.0033Å$^{-1}$ to 0.56Å$^{-1}$. The fitting results of [10] seem close to [8] although the extended range of q led to the increasing of a membrane thickness approximately on 5Å and decreasing of the number of water molecules in the hydrophilic region from 5.7±0.2 in [8] to 3.90±0.03 in [10].

In the present paper, we apply more realistic and complicated model to evaluate the internal DMPC membrane structure. Firstly, the distribution of water molecules through the membrane bilayer was described as sigmoid function. The linear distribution of water in region of polar head group was proposed by J. Nagle [11]. The linear functions with discontinuity point on the boundary between the membrane and the bulk water were used in [12] to fit SANS curve by Kratky-Porod method. Physically, boundary conditions for water distribution function should be formulated as necessity to have a continuous first derivative. Linear functions are not satisfied to this request. We selected the sigmoid function as a reasonable function that can satisfy to the boundary conditions. This choice is supported by the water distribution function calculated via computer simulations [14] of DPPC fluid phase. As shown in [13], the simulated water distribution function is more similar to the sigmoid in comparison with the linear.

Secondly, the neutron scattering density of polar head group was modeled by the Gauss function. The Gauss functions were successfully used by Wiener and White to restore the internal membrane structure from neutron and X-ray diffraction experiment [15]. Later, this approach was further developed for the interpretation of X-ray reflectivity from phospholipid monolayer on water surface [16]. It was shown that strip-functions can be correctly applied to describe the reflectivity curve only for q<0.5Å$^{-1}$. The Gauss functions are necessary to be introduced to describe experimentally measured reflectivity curve for q>0.5Å$^{-1}$ [17]. This approach has important physical sense, because the molecular groups in phospholipid bilayer (monolayer) are fluctuated near it equilibrium position. Any group can be described via two parameters: average group position and the Gauss distribution function around it. According to this approach, the geometrical sense of lipid membrane thickness changes to the distribution function of polar head group position.

## 1. Experiment

Unilamellar dimyristoylphosphatidylcholine vesicles (DMPC) were prepared by extrusion of 15mM (1% w/w) suspension of DMPC in $D_2O$ through filters with a pore diameter of 500Å.

The SANS spectra from the unilamellar vesicle population at T=30°C were collected at the SANS spectrometer of the Swiss Spallation Neutron Source at the Paul Scherrer Institute (PSI), Switzerland. Three sample-to-detector distances were used: 2m, 6m, and 20m. Neutron wavelength was 4.7±0.47Å. The spectra were normalized on the macroscopic cross-section of $H_2O$. The value of the incoherent background was a fitted parameter in the model calculations.

Note that the accuracy of the vesicle structure fitting strongly depends on the experimentally measured range of the scattering vector [10]. This means that the restored parameters of the internal membrane structure depend on the value of maximum q measured experimentally. For the system under study, the experimental conditions allow to collect a scattering curve in the q range from $q_{min}$=0.0033Å$^{-1}$ to $q_{max}$=0.56Å$^{-1}$.

## 2. Formulation of the fitting problem (the SFF model)

The macroscopic coherent scattering of monodispersed population of vesicles is defined by the formula [18]:

$$\frac{d\Sigma}{d\Omega}_{mon}(q) = n \cdot A^2(q) \cdot S(q) \quad (1)$$

where $n$ is a number of vesicles per unit volume, $A(q)$ is the scattering amplitude of vesicle, $S(q)$ is the vesicle structure factor that is calculated as in [7]; $q$ is the length of scattering vector ($q = 4\pi \sin(\theta/2)/\lambda$, $\theta$ - the scattering angle, $\lambda$ - the neutron wavelength).

The scattering amplitude in the spherically symmetric case is equal [18] to

$$A(q) = 4\pi \cdot \int \rho(r) \cdot \frac{Sin(qr)}{qr} \cdot r^2 \cdot dr \quad (2)$$

Here $\rho = \rho_C - \rho_{D2O}$ is the neutron contrast between the scattering length density of the lipid bilayer $\rho_C$ and $D_2O$ ($\rho_{D2O}$=6.4·10$^{10}$ cm$^{-2}$).

Eq.(2) can be rewritten as follows [6]:

$$A(q) = 4\pi \cdot \int_{-d/2}^{d/2} \rho(x) \cdot \frac{Sin[(R+x) \cdot q]}{(R+x) \cdot q} \cdot (R+x)^2 \cdot dx \quad (3)$$

Here $R$ is the radius of vesicle, $d$ is the membrane thickness. Integration of eq.(3) in assumption $R >> d/2$, $R+x \approx R$ gives

$$A_{sff}(q) = 4\pi \cdot \frac{R^2}{qR} \cdot Sin(qR) \cdot \int_{-d/2}^{d/2} \rho(x) \cdot Cos(qx) \cdot dx \quad (4)$$

Thus, the macroscopic cross-section of the monodispersed population of vesicles (in case, $S(q)$=1) can be written as

$$\frac{d\Sigma}{d\Omega}_{mon}(q) = n \cdot F_s(q,R) \cdot F_b(q,d) \tag{5}$$

where $F_s(q,R)$ is a form-factor of the infinitely thin sphere with radius $R$ [19]

$$F_s(q,R) = \left(4\pi \cdot \frac{R^2}{qR} \cdot Sin(qR)\right)^2 \tag{6}$$

and $F_b(q,d)$ is a form-factor of the symmetric lipid bilayer

$$F_b(q,d) = \left(\int_{-d/2}^{d/2} \rho(x) \cdot Cos(qx) \cdot dx\right)^2 \tag{7}$$

Eqs.(5)-(7) present the separated form-factor model (SFF) for large unilamellar vesicles [6]. This model has an advantage due to the possibility of describing the membrane structure via representation of $\rho(x)$ as any integrable function.

The vesicle polydispersity is described by the nonsymmetric Schulz distribution [2,20]

$$G(R) = \frac{R^m}{m!} \cdot \left(\frac{m+1}{\overline{R}}\right)^{m+1} \cdot \exp\left[-\frac{(m+1) \cdot R}{\overline{R}}\right] \tag{8}$$

where $\overline{R}$ is an average vesicle radius, $m$ is a coefficient of polydispersity. Relative standard deviation of vesicle radius is $\sigma = \sqrt{\frac{1}{(m+1)}}$.

Thus, macroscopic cross section $d\Sigma(q)/d\Omega$ has the following form:

$$\frac{d\Sigma}{d\Omega}(q) = \frac{\int_{R\min}^{R\max} \frac{d\Sigma}{d\Omega}_{mon}(q,R) \cdot G(R,\overline{R}) \cdot dR}{\int_{R\min}^{R\max} G(R,\overline{R}) \cdot dR} \tag{9}$$

where $R_{min}$=100 Å, $R_{max}$=1000 Å.

The experimentally measured cross section $I(q)$ is not completely equal to the actual macroscopic cross-section $I_m(q)$ because the resolution function of the spectrometer is not a delta-function. The experimental cross-section $I(q)$ can be given by

$$I(q) = I_m(q) + \frac{1}{2} \cdot \Delta^2 \cdot \frac{d^2 I_m(q)}{dq^2} \tag{10}$$

where $\Delta^2$ is a second moment of the resolution function [21], $I_m(q)=d\Sigma(q)/d\Omega$ (see eq.(9)). In general, the calculation of the resolution function is a separate problem requiring a special study; we do not consider it in this paper. $\Delta=0.1q$ was used for $L_{sd}$=2m and 6m, and $\Delta=0.2q$ was used for $L_{sd}$=20m.

For the fitting of experimental data we used function $\chi^2$:

$$\chi^2 = \frac{\sum_{i=1}^{N}[(\Sigma_{exp}(q_i) - I(q_i))/\delta(q_i)]^2}{N-k}$$
(11)

where $\delta$ are the experimental statistical errors, $N$ is a number of experiment points, $k$ is a number of unknown parameters, $\Sigma_{exp}$ – experimentally measured coherent macroscopic cross sections.

The fitting parameters are: average vesicle radius $\overline{R}$, coefficient of polydispersity $m$, and parameters of function $\rho_\Gamma(x)$ modeling the neutron scattering length density. We consider the incoherent background as another unknown parameter of the model.

Number of vesicles per unit volume $n$ can be obtained in the following way. It is known that the volume of molecular DMPC in the liquid phase is $V_{DMPC}=V_0+n_W V_{D2O}$ (where $V_0=1101 Å^3$ - volume of 'dry' DMPC molecule [13], $n_W$ - number of water molecules per one DMPC molecule, $V_{D2O}=30$ $Å^3$ - volume of the water molecule). Hence, the volume of the lipid bilayer can be calculated by formula

$$V = 4\pi/3 \ [(\overline{R}+d/2)^3 - (\overline{R}-d/2)^3].$$
(12)

Here, $d$ is a membrane thickness. So, $M=V/V_{DMPC}$ is the number of DMPC molecules in a single vesicle. The number of DMPC molecules in $cm^3$ can be estimated as $C=89.17\cdot 10^{17}$. It means that we can put $n=C/M$.

Alternative way to estimate number $n$ is to put $M=S/A$ where $S=8\pi \overline{R}^2$ is a total area of the vesicle surface, $A$ is a surface area of the DMPC molecule. $A$ can be defined from formula $A=2 V_{DMPC}/d$.

Note that $n$ is not constant, it depends on unknown fitting parameters $\overline{R}$ and $d$. In both ways we have a problem of unknown parameter $n_W$ (number of the water molecules). In [9] $n_W$ was fixed; in the present fitting we approximately put $V_{DMPC} \sim V_0$; number $n_W$ is estimated after all other vesicle parameters have been calculated. This simplification can only decrease an accuracy in the calculation of average vesicle number in unit volume that is beyond our interest; it cannot have influence on the function $\rho(x)$.

Finally, in order to estimate the fit quality, we used the following formula:

$$R_I = \frac{1}{N} \cdot \sum_{i=1}^{N} \left( \frac{|\frac{d\Sigma}{d\Omega}(q_i)| - |\frac{d\Sigma}{d\Omega}_{exp}(q_i)|}{|\frac{d\Sigma}{d\Omega}_{exp}(q_i)|} \right)^2 \times 100\%.$$
(13)

### 3. <u>Results and discussion</u>

To fit the SANS data in the framework of SFF model, the Fortran code was developed using the minimizing code DFUMIL from the JINRLIB library (JINR, Dubna).

Internal structure of the lipid bilayer was simulated by two types of function $\rho(x)$. We suggested that $\rho(x)$ is superposition of functions $\rho_{ph}(x)$ (scattering length density of 'dry' lipid) and $\rho_w(x)$ - the water distribution inside the vesicle. We used the strip ('step') and the Gauss function to model the 'dry' lipid. Hydration of vesicle was simulated by sigmoid function (dashed lines in Fig. 1). The distribution shown ine Fig. 1a, is defined by the following formulas:

$$\rho_c(x) = \rho_{ph}(x) + \rho_{ch}(x) + \rho_w(x) \tag{14}$$

$$\rho_{ph}(x) = \begin{cases} l_{ph} \cdot 2/[(d-D) \cdot A], & D/2 < |x| < d/2 \\ 0, & -D/2 < x < D/2 \end{cases} \tag{15}$$

$$\rho_{ch}(x) = \begin{cases} l_{ch} \cdot 2/[D \cdot A], & -D/2 < x < D/2 \\ 0, & D/2 < |x| < d/2 \end{cases} \tag{16}$$

$$\rho_w(x) = \frac{\rho_{D_2O}}{1 + \exp(\frac{x_W - x}{\sigma_w})} \tag{17}$$

where $l_{ph} = 6.008 \cdot 10^{-12}$ cm, $l_{ch} = -3.24 \cdot 10^{-12}$ cm, $\rho_{D_2O} = 6.4 \cdot 10^{10}$ cm$^{-2}$, $A = 2V_{DMPC}/d$, $A$ – the surface area of the DMPC molecule, $D$ – thickness of hydrocarbon chain region. Fitting parameters are: $d$, $D$, $x_W$, $\sigma_W$, $m$, $\overline{R}$, IB (incoherent background).

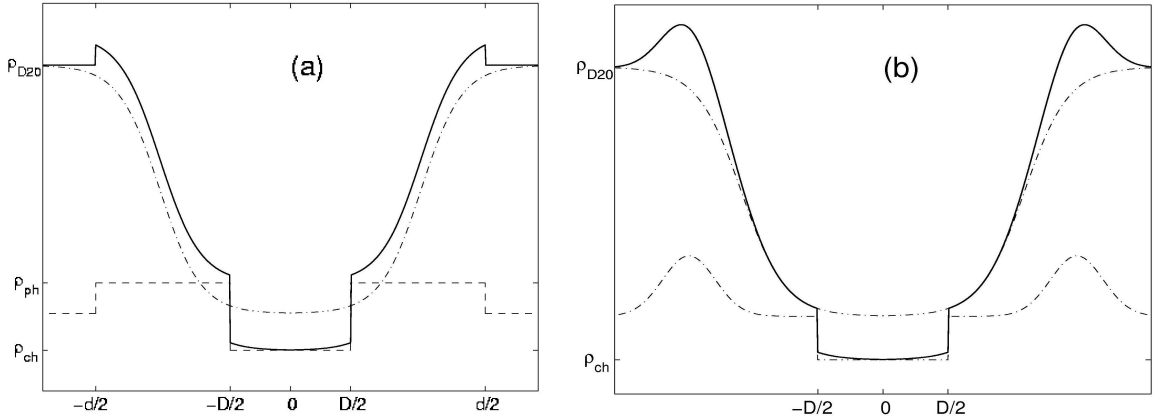

*Fig. 1. Two models of the neutron scattering length density across the bilayer. The dashed lines demarcate the length scattering density of the 'dry' lipid bilayer: (a) – the strip-function, (b) – the Gauss function (fluctuation model). Water distribution is simulated by sigmoid function (the dash-dots on the both figures (a) and (b)). The solid lines show the total neutron scattering length density across the bilayer.*

For the Gauss distribution (Fig.1b) we have

$$\rho_{ph}(x) = f(x) \cdot l_{ph}/A, \tag{18}$$

$$f(x) = \frac{1}{\sqrt{2\pi}\Sigma} \exp\left(-\frac{(x-x_0)^2}{2\Sigma^2}\right), \quad (19)$$

$\rho_{ch}$ and $\rho_W$ are determined from the above eqs. (15),(16); $d$ is defined in this case as $d=2x_0$ and, consequently, $A$ is defined as $A=V_{DMPC}/x_0$. Fitting parameters in this case are: $D$, $x_0$, $x_W$, $\sigma_W$, $\Sigma$, $m$, $\overline{R}$, IB.

Fitting results for the spectra of the PSI SANS experiment are presented in Fig. 2 and Table 1. In this calculation, the structure factor was included as in [7]. For the resolution function correction, we used the value $\Delta q/q=20\%$ at small q and $\Delta q/q=10\%$ at large q. In both cases we obtained $m=12.5\pm0.1$ (polydispersity 27%); it is in the agreement with our previous results [8] and [10].

Table 1. *Parameters of the DMPC vesicles ($T=30\ ^oC$) calculated in the framework of SFF model for two forms of the scattering length density of neutrons across the lipid bilayer (see Fig. 1a,b).*

|     | $\overline{R}$, Å | $d$, D, Å | $x_W$, $\sigma_W$ | $X_0$, $\Sigma$ | IB, $10^{-3}$cm$^{-1}$ | $R_I$, % |
|-----|-------|-----------|---------|--------|------------------------|----------|
| (a) | 273.5±0.4 | 55.2±0.9 | 18.46±0.04 |  | 6.06±0.01 |  |
|     |           | 17.0±1.5 | 2.89±0.05  |  |           | 0.094 |
| (b) | 272.3±0.4 |          | 18.3±0.6   | 25.3±0.4 | 6.05±0.02 |  |
|     |           | 21.4±2.6 | 6.91±0.2   | 3.43±0.7 |           | 0.095 |

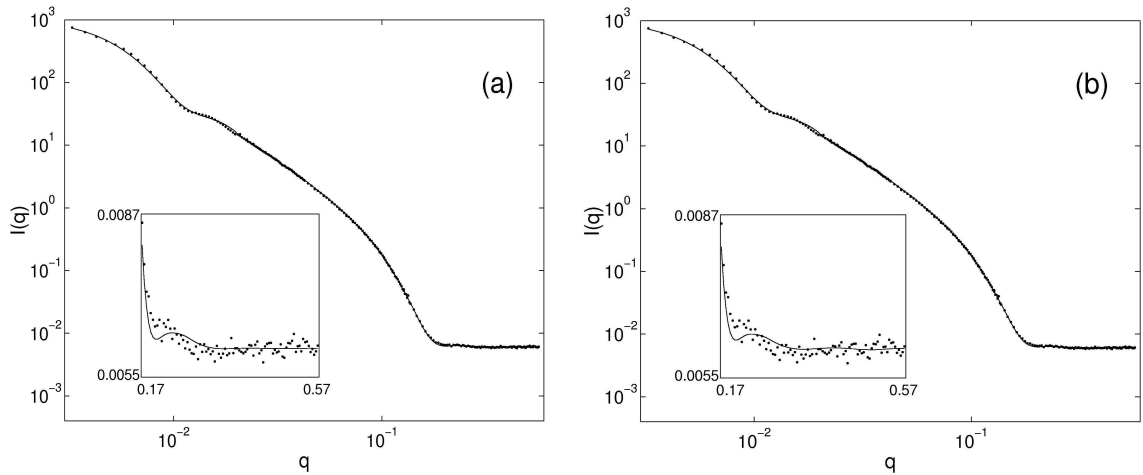

Fig. 2 *Fitting results of the DMPC vesicle spectrum for two models of the internal structure of the lipid bilayer given in Fig. 1 a,b. The points are the experimental data, the solid lines show results of the fitting. The insets show in detail the curves for large q.*

It is seen from the Table 1 that the involving of the sigmoid water distribution notably increased the membrane thickness. In the case (a) the calculated membrane thickness is unreasonably large $d=55.2$Å. For the case (b), the thickness of the membrane can be estimated as $2 \cdot x_0 = 50.6 \pm 0.8$; it is in better agreement with our result [10] for the linear water distribution from the PSI SANS spectrum, $d=47.4\pm0.04$Å. The linear water distribution gives $d=42.7\pm0.42$Å from the YuMO spectrum [8]. Kratky-Porod model give $d=44.5\pm0.3$ in [9], and $d=48.08\pm0.45$Å in [12]. J. Nagle's estimation is $d=44.2$Å [13].

The number of water molecules per one DMPC molecule can be calculated as

$$n_w = \frac{A}{l_{D2O}} \cdot \int_0^{d/2} \rho_w(x)dx. \qquad (20)$$

Here we used $A=2V_0/d= 43.5$Å$^2$. This value of A is near the value $44.8\pm1.8$Å$^2$ obtained from the wide-angle X-ray diffraction from DMPC multilamellar vesicles with 50% hydration [22] and can be used as a first-order approximation. Both models give almost the same values $n_W$: $12.5\pm1.1$ in the case (a) and $12.6\pm1.1$ in the case (b). However, the number of water molecules in the region of hydrocarbon chains (integration in equation (20) from 0 to $D/2$) is sufficiently different and corresponds to the value 0.11 in the case (a), and to 2.2 in (b). This means that the model (a) shows a small penetration of water molecules inside the region of hydrocarbon chains. A standard deviation of the water distribution function in the case (a) is 2.4 times smaller than in the case (b). Consequently, the probability to find water molecules in the region of hydrocarbon chains is sufficiently larger for the case of fluctuated model (b).

It is well known from the neutron diffraction that water molecules easy penetrate through bilayer [23]. Our SANS results on the basis of the fluctuated model (b) clearly support this experimental fact via calculation of water distribution function across the bilayer. Taking into account the X-ray time-resolved diffraction results about 3 min water diffusion through membrane under ice induced dehydration [24], one can expect that fluctuated model of lipid bilayer gives more realistic results.

In our calculations, we used estimation $A=43.5$Å$^2$, but in reality this value corresponds to the area of membrane surface occupied by a dry DMPC molecule. Now one can make a correction of the DMPC surface area, taking into account that water molecules increase the volume occupied by one DMPC molecule. The corrected value: $V_{DMPC} = V_0 + n_W \cdot 30$Å$^3 = 1480$Å$^3$. Hence, taking into account the DMPC "swelling" by water, we obtain $A_w = 2 \cdot V_{DMPC}/d = 58.5\pm2$Å$^2$. This value $A$ is close to the value $59.6$Å$^2$ published by J. Nagle [13], value $58.9\pm0.8$Å$^2$ in [9], and value $58.8\pm0.5$Å$^2$ in [12].

The larger value of the membrane thickness in the present study gives approximately two times larger value of water molecules $n_w=12.6$ per one DMPC molecule (relative to the previously published results $n_w=7.2$ [13] and 5.7 [8]). One can check the correctness of the evaluation of water distribution function via calculation of the value of $n_w$ from eq.(20) with value of $A=59.6$Å$^2$ and $d=44.2$Å according to [13]. This gives 6.4 as a number of water molecules per one DMPC molecule. The obtained result is intermediate between the previously published values. Thus, water distribution function looks quite correct in the region of polar head groups.

Full width at half height of the distribution function of polar head groups $2.36 \cdot \Sigma = 8.0\pm1.7$Å can be used to estimate the region of polar head group location. The obtained value of polar head group location is near the 9Å thickness of polar head groups obtained from the X-ray diffraction [13].

The thickness of hydrocarbon chains $D=21.4\pm2.6$Å obtained in the present paper is not at contradiction with Nagle's result, 26.2 Å obtained via the strip-function approximation of electron density profile [13]. This approach describes quite precisely the hydrophobic/hydrophilic boundary, as was shown in [5], but strong definition of this boundary as exact value is far beyond the reality. The hydrophobic/hydrophilic boundary is some broad region near the value $x_W-\sigma_W=11.4\pm0.8$Å that reasonably corresponds to the end of hydrocarbon chains $D/2=10.7\pm1.3$Å. It is important to note that three parameters: $x_W$, $\sigma_W$, and $D$ were independent in our calculations. The description of water distribution function in the region of hydrophobic/hydrophilic boundary and the thickness of hydrocarbon chains are not in variance with each other.

An important result of our study is the explanation of the earlier published differences in the evaluation of membrane thickness for the case $\rho(x)=$const and $\rho(x)$ as step-function. The calculation of the DMPC membrane thickness on the basis of $\rho(x)=$const model gives value of membrane thickness $36.7\pm0.1$Å, which increases up to the value of $42.1\pm0.4$Å at the application of strip-function model of $\rho(x)$ [8]. The total neutron scattering length density across the bilayer obtained in our study (see Fig. 1b) clearly demonstrate; that modeling of $\rho(x)$ via box model with $\rho(x)=$const inside the bilayer leads to decreasing of membrane thickness to the value about 34Å. This underestimation of the membrane thickness on the value of 5.4Å is the result of deep $D_2O$ penetration inside the bilayer.

### 4. Conclusions

On the basis of the SFF model, the scheme and code of fitting the SANS spectra of polydispersed vesicle population have been developed taking into account: structure factor, spectrometer resolution function, and internal structure of vesicles. The fluctuated model is proposed to describe the internal structure of phospholipid bilayer. This model describes reasonably well the region of hydrocarbon chains, thickness of polar head groups and water distribution function across the bilayer. The penetration of water molecules through the bilayer is proved via direct calculation of water distribution function across the bilayer.

### Acknowledgements

We are grateful to I.V. Puzynin, V.L. Aksenov, N. Popa and I.V. Amirkhanov for useful discussions. The investigation was supported by the Ministry of Science and Technology of the Russian Federation (grant N°40.012.1.1.1148), Grant of Leading Scientific School, and RFBR (grant N°03-01-00657).